# Circulating Lymphocytes Preservation in Lung Cancer Stereotactic Body Radiation Therapy with Ultra-Fast Proton Delivery Using Modularized Pin Ridge Filters

Duncan Bohannon, Ahmal Jawad Zafar, Sibo Tian, Williams Stokes, Zachary S. Buchwald, Sunil W. Dutta, William A. LePain, Zachary Diamond, Anees Dhabaan, Hania Al-Hallaq, Xiaofeng Yang, and Jun Zhou

Purpose:
Radiation-induced lymphopenia is an increasingly recognized toxicity in lung radiotherapy and has been linked to radiation exposure to circulating lymphocytes (CL). In intensity-modulated proton therapy (IMPT), prolonged pencil beam scanning (PBS) delivery may increase CL dose. We recently developed a patient-specific pin ridge filter (pRF) framework that enables ultra-fast proton delivery with a single beam energy. This study evaluated whether pRF-based lung stereotactic body radiotherapy (SBRT) plans delivered at conventional (pRFCONV) and FLASH dose rates (pRFFLASH) improve immune sparing using time-resolved blood dose accumulation and CL survival modeling.
Methods:
pRF plans were created for 10 lung SBRT patients previously treated with IMPT. PBS delivery simulations modeled spot delivery, scanning, and energy switching. Blood dose-volume histograms (bDVHs) were calculated with the hematological dose framework. CL survival fractions (SF) were estimated from bDVHs with saturation and linear-quadratic models derived from in-vitro survival data for CD4/CD8 CL.
Results:
Compared with IMPT, pRFCONV/pRFFLASH plans reduced delivery time (mean reductions: 85.3/99.9%) and irradiated blood volume per fraction (mean reductions: 52.9/81.3%). pRFCONV/pRFFLASH plans reduced blood V5cGy by 26.4/39.4%, and V50cGy by 4.5/6.9%, respectively. pRF plans improved modeled CL survival across all models and subpopulations. Unstimulated CD4/CD8 CL had the largest SF differences, for which mean saturation-model SF improved by 7.9/8.6% for pRFCONV ($p=0.03/0.02$) and 9.6/10.4% for pRFFLASH ($p=0.02/0.01$), respectively.
Conclusion:
pRF plans improved modeled CL survival by significantly shortening delivery time and reducing irradiation of circulating blood. Our findings suggest that pRF's ultra-fast delivery may provide a practical strategy for immune sparing in proton lung SBRT.

# 1. Introduction

Radiation-induced lymphopenia (RIL), a condition in which lymphocytes in the blood are significantly depleted following radiation exposure, is a prevalent and clinically significant toxicity in radiotherapy [1-3] that leads to worsened survival, recurrence, and metastasis [4]. RIL has been correlated with the dose to circulating lymphocytes (CL) [5-10], suggesting that it is frequently caused by radiotherapy due to the large volume of blood irradiated during treatment [11] combined with the high radiosensitivity of CLs [12-15].

Treatment modality, site, and fractionation strongly influence RIL risk. RIL occurs particularly often in thoracic patients where larger volumes of blood are irradiated [16,17]. Compared with intensity-modulated radiation therapy (IMRT), proton therapy has been associated with reduced rates of severe RIL. In esophageal cancer cohorts, studies have found that proton therapy decreases the rate of grade 4 RIL by 23% to 70% [7,18]. This difference has been attributed to proton therapy's lowered integral dose, which can reduce the irradiated blood volume [19]. Hypofractionated techniques such as stereotactic body radiation therapy (SBRT) may further reduce cumulative circulating blood exposure by decreasing the number of fractions and has been associated with improved lymphocyte preservation in selected disease sites [6, 16, 20-23]. These findings suggest that CL survival is strongly influenced by the volume of blood exposed to low doses outside the target.

Proton FLASH pencil beam scanning (PBS) radiotherapy, characterized by ultra-high dose rates ($\geq$40 Gy $s^{-1}$), has been computationally shown to reduce CL exposure compared with conventional intensity-modulated proton therapy (IMPT), largely by significantly decreasing the volume of irradiated blood during treatment [24]. Pin-ridge filters (pRFs), introduced in the FLASH radiotherapy literature [25], enable delivery with single-energy proton beams that achieve the ultra-high dose rates required to trigger the FLASH effect while preserving plan quality [26,27]. We recently developed a framework for streamlined patient-specific pRF design that enables the generation of clinically deliverable pRF treatment plans [28,29]. Energy-layer switching (ELS) is a significant contributor to PBS delivery time. Except for some compact systems that achieve 0.05s ELS time [30], most commercial proton centers use much slower systems with switching durations ranging from 0.2 to 2 seconds [31-33]. pRFs eliminate ELS times by using a single beam energy. As a result, even when delivered at non-FLASH conventional dose rates, pRF plans can substantially reduce overall treatment time compared with IMPT. Zafar et al. reported that pRF liver plans delivered at conventional dose rates reduced median delivery time by 83.4% compared to IMPT [29]. As a result, pRF's fast delivery could decrease CL exposure, making it a promising technique to enhance immune preservation in proton therapy. This study aimed to investigate whether clinically deliverable pRF plans generated from our framework can improve immune preservation compared with traditional IMPT for lung SBRT patients, a population particularly vulnerable to RIL [16,17].

Early studies evaluated immune preservation by using mean organ doses to calculate effective dose to circulating immune cells (EDIC) [34]. Subsequent studies improved upon the EDIC framework by developing models that estimate the circulating blood dose as a surrogate for CL dose. This work utilizes Hematological Dose (HEDOS) [35], an open-source stochastic blood-flow framework for modeling blood circulation and calculating blood dose-volume histograms (bDVHs) that has been used to predict RIL [36] and investigate the impacts of varying treatment modality [35, 36], plan parameters [37], and treatment site [38] on immune preservation. This study enhanced the modeling of bDVHs by introducing a modified dose accumulation framework that incorporated a detailed simulation of PBS delivery comprised of spot delivery, scanning, and ELS timing parameters from our institution.

This work retrospectively compared bDVHs from a cohort of 10 lung SBRT patients previously treated at our institution across three planning strategies: (1) clinical IMPT plans, (2) pRF plans delivered at conventional dose rates, and (3) the same pRF plans delivered at FLASH dose rates. The clinical impact of observed bDVH differences was assessed by estimating CL survival across multiple lymphocyte subpopulations with a traditional linear-quadratic (LQ) lymphocyte survival model from Nakamura et al. [39] and a recently proposed saturation survival model from Pham et al. [13] that are both based on in vitro lymphocyte data. Although these survival models don't explicitly represent the in vivo dose response of CL, this work applied them as comparative response functions to estimate relative CL injury. We hypothesize that pRF's ultra-fast delivery will enhance CL survival, providing a practical approach to mitigate RIL and improve immune preservation in proton lung SBRT.

# 2. Methods

## 2.1 pRF Plan Creation

pRF plans were generated in RayStation 2023B, RaySearch Laboratories, Stockholm, Sweden. using an inverse-planning framework originally described in [28,29] that jointly optimizes pRF geometry and spot weights. Figure 1 illustrates the four-stage pRF planning process for an example lung patient.

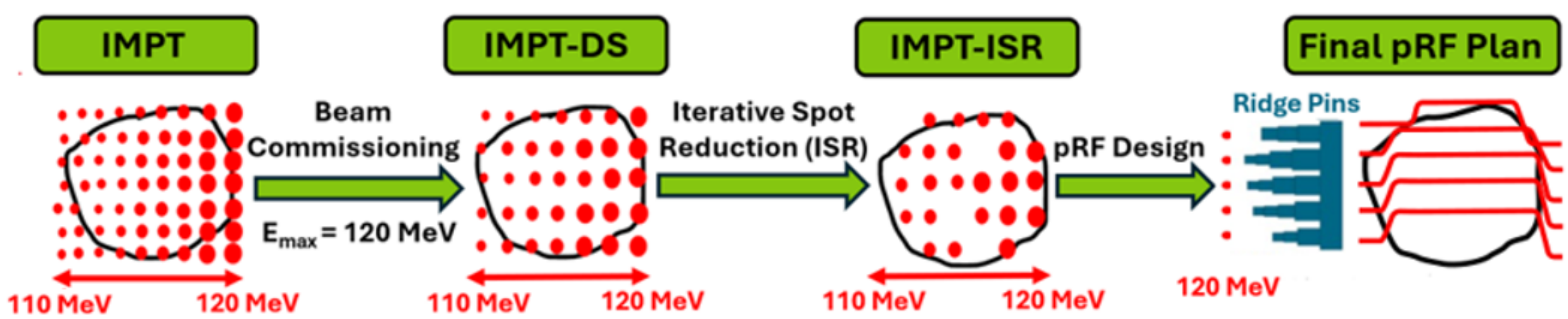

**Figure 1:** Diagram of pRF planning workflow for an example lung patient. Red dots represent spots; dot size increases with spot weight, and columns of dots represent energy layers. The planning process starts with a conventional IMPT plan. For this example, a beam model with a maximum energy matched to the IMPT plan (120 MeV) is commissioned. A downstream plan (IMPT-DS) is then created by reoptimizing the IMPT plan using the newly commissioned beam. An iterative spot reduction (ISR) algorithm is applied to the IMPT-DS plan to remove low-weight spots, creating an IMPT-ISR plan. The IMPT-ISR plan is used to design the pRF geometry for each beam. Finally, the IMPT-ISR is converted into a monoenergetic plan modulated by pRFs and reoptimized to create the final pRF plan.

First, a conventional IMPT plan is created. Next, a beam model is commissioned within the TPS with parameters tuned for pRF design. The maximum energy of the commissioned beam is matched to the maximum energy required by the IMPT plan, preserving distal target coverage when using a single beam energy. To avoid beam commissioning for each patient, monoenergetic beams spanning 110–140 MeV in 10 MeV steps were commissioned. For each patient, the commissioned beam with energy closest to $E_{max}$ was selected. pRFs are modeled as water-equivalent structures consisting of pyramidal ridge pins mounted on a base and represented by stacked water-equivalent thickness (WET) slabs of 5 mm thickness. The commissioned beam's energy layers correspond to the residual energies after transmission through different stacks of ridge pins.

The commissioned beam is used to create a downstream plan (IMPT-DS). IMPT-DS optimization remains multi-energy but is constrained to the energy layers available in the commissioned beam. An iterative spot reduction (ISR) algorithm described in [28,29] is then applied to the IMPT-DS plan to create an IMPT-ISR plan. ISR removes low-weighted spots, thereby reducing the complexity of the final pRF geometry.

Spot energies and weights from IMPT-ISR are converted into pRF geometry using the process detailed by Ma et al. [28]. Each ridge pin is aligned with a pencil beam direction (PBD) from the IMPT-ISR plan. Modulation of the $E_{max}$ monoenergetic beam is achieved by varying the surface areas of each ridge pin's steps. For a ridge pin step $i$ at PBD $p$ with WET of $t_{p,i}$, the corresponding $i$th Bragg curve's range (80% falloff), $R_{p,i}$, for protons passing through the step is given by:

$$R_{p,i} = R_0 - t_{p,i} \tag{1}$$

where $R_0$ is the range of the monoenergetic beam. Let $s_{p,i}$ represent the side length of the square surface of step $i$. The relative weight of Bragg curve $i$, $\overline{w}_{p,i}$, is recursively calculated with Equations 2-3 given the width of the bottom step, $s_{p,1}$:

$$\overline{w}_{p,i-1} = \frac{s^2{}_{p,i+1} - s^2{}_{p,i}}{s^2{}_{p,i}} \tag{2}$$

$$\overline{w}_{p,i-1} = \frac{w_{p,i}}{\sum_{i=1}^{N} w_{p,i}} \tag{3}$$

where $w_{p,i}$ is the relative weight of spot $i$ in the IMPT-ISR plan, and $N$ is the number of energies/spots within the corresponding PBD. Ma et al. determined that the selected value of $s_{p,1} = 0.6cm$ was an optimal value for beam modulation quality [28].

## 2.2 Blood Circulation Simulation

This study utilized the HEDOS framework to simulate blood circulation. HEDOS [35] implements a compartmental blood-flow model based on ICRP 89 [40], which represents the body as 28 interconnected compartments. Each compartment is assigned a fraction of the total blood volume and exchange flow rates with connected compartments.

HEDOS initializes the simulation by dividing the total blood volume into individual blood particles (BPs) and assigning them to each compartment according to ICRP reference conditions. During the simulation, BPs stochastically transition between compartments according to probabilities derived from the intercompartmental flow rates. The resulting BP trajectories provide a time-resolved compartmental distribution of circulating blood throughout the body. In this study, all HEDOS simulations used 100,000 BPs to minimize stochastic fluctuations in the blood dose histogram and a time step resolution of 50ms, the finest time resolution available [35]. For male/female patients, total blood volume and cardiac output were set to ICRP 89 reference values of 5.3/3.9 L and 6.6/5.9 L min$^{-1}$, respectively [40].

## 2.3 Temporal Simulation of PBS

To accurately represent the time-dependent nature of radiation delivery, temporal simulations of PBS were performed. PBS delivery was divided into three components: spot delivery times ($T_d$), scanning times between spots ($T_s$), and ELS ($T_e$):

$$T_d = \left[T_{d_o}, T_{d_1}, \dots, T_{d_n}\right]$$
$$T_s = \left[T_{s_o}, T_{s_1}, \dots, T_{s_{n-1}}\right]$$
$$T_e = \left[T_{e_o}, T_{e_1}, \dots, T_{e_{l-1}}\right]$$

Where *n* is the number of spots and *l* is the number of energy layers. These components were calculated using a clinically validated beam time script previously developed at our institution that describes the timing structure of our Varian ProBeam system. The script was developed using more than 10,000 treatment log files collected over one month together with experimental measurements of maximum dose rates, energy switching times, and scanning magnet speeds. Because scanning speeds and ELS times vary during clinical PBS delivery, average values were used to reduce variability in temporal simulations.

### 2.3.1 Spot Delivery

During PBS delivery, the dose rates for each energy layer are determined by its smallest weighted spot and is limited by the maximum cyclotron output. Our Varian ProBeam system has an average minimum spot time of 3.14 ms, giving the following relation for the energy layer dose rate:

$$\dot{D}_j = min \begin{cases} \dfrac{(MU_{min})_j}{3.14 \times 10^{-3}\ s} \\ (\dot{D}_{max})_j \end{cases} \quad (4)$$

where $(MU_{\min})_j$ is the minimum spot MU within energy layer $j$ and $\dot{D}_{max}$ is the corresponding maximum dose rate. During the development of the clinical beam time script, $\dot{D}_{max}$ was measured at 70, 150, and 242 MeV. Values at other energies were determined with linear interpolation.

A different approach was used to calculate layer dose rates for FLASH delivery simulations. To achieve FLASH dose rates, higher nozzle currents are required. This study simulates FLASH dose rates using a 500nA current, which is close to the maximum nozzle current achievable for our Varian ProBeam system. This nozzle current is theoretical because it can only be achieved by the highest energy beam. It is used to evaluate the potential additional CL sparing afforded by FLASH delivery. Layer dose rate for FLASH simulations was calculated as:

$$\dot{D}_j = \frac{\left(6.24 \times 10^9 \frac{Protons}{nC}\right) \times 500 \ \frac{nC}{s}}{I(E) \frac{Protons}{MU}} = \frac{3.12 \times 10^{12}}{I(E)} \quad \frac{MU}{s} \tag{5}$$

where $I(E)$ are institutional proton per MU calibration values defined from 70-240 MeV. Once $\dot{D}_j$ is calculated, delivery times for the spots within layer *j* was determined as:

$$T_{d_i} = \frac{MU_i}{\dot{D}_j} \tag{6}$$

where $MU_i$ is the MU of spot i, and $\dot{D}_j$ is the dose rate of energy layer j.

### 2.3.2 Scanning

Varian ProBeam systems simultaneously operate X and Y scanning magnets to steer the pencil beam between spots. Thus, scanning time between spots is determined by the magnet with the longest travel time and was calculated as:

$$T_{s_i} = \max\left(\frac{|\Delta x_i|}{v_x}, \frac{|\Delta y_i|}{v_y}\right) [ms] \tag{7}$$

Where $T_{s_i}$ is the scanning time between spot i and i+1, $\Delta x_i$ and $\Delta y_i$ are the spatial displacements between the adjacent spots in the X and Y directions, and $v_x$ and $v_y$ are the corresponding scan speeds. Average scanning speeds of $v_x = 0.43$ cm/ms and $v_y = 0.85$ cm/ms were used.

### 2.3.3 ELS

Energy switching times were measured during the development of the clinical beam time script over energy differences of 0-90 MeV. ELS times followed a linear relationship:

$$T_{e_j} = 789ms + 122 \frac{ms}{MeV} \Delta E_j \tag{8}$$

Where $\Delta E_j$ is the energy difference between layers $j$ and $j + 1$. The slope of this linear fit represents the average time added from energy degrader motion per MeV, and the intercept represents the average cyclotron preparation time.

## 2.4 Blood Dose Accumulation

A blood dose accumulation framework was developed to synchronize the spatiotemporal distribution of BPs from HEDOS with the temporal PBS delivery simulation. Figure 2 illustrates the blood dose accumulation process for a single PBS beam. First, a HEDOS simulation is generated. Then, the PBS delivery timeline is discretized into $m$ HEDOS timesteps of duration $\Delta t$. The beam dose, $D$, is divided into partial doses, $D_i$, which only include spots delivered during their corresponding timestep. Partial doses are then used to accumulate dose to BPs located within the organ of interest at each timestep.

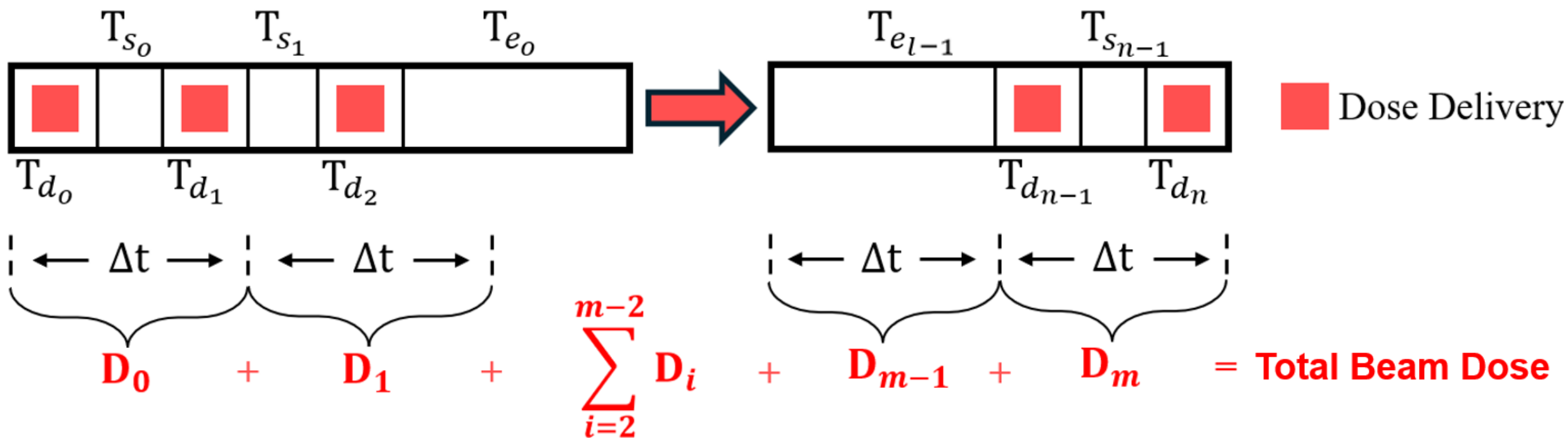

**Figure 2:** Diagram of PBS beam delivery synchronized with HEDOS. The PBS delivery timeline is comprised of $n$ spot delivery times ($T_d$), $n$-1 scanning times between spots ($T_s$), and $l$-1 energy switching between $l$ energy layers ($T_e$). The HEDOS model provides the positions of BPs within the body for $m$ time steps. The total beam dose, D, is divided into partial doses at each timestep and added to BPs.

Because HEDOS only provides BP compartment locations, BP positions inside the lung segmentation were determined with a random walk to approximate intrapulmonary circulation. The initial positions of BPs within the segmentation are randomly sampled according to a uniform distribution ($U$), and at subsequent timesteps, their positions are updated with the equations:

$$r_{i+1} = r_i + v_i \Delta t \qquad (9a)$$

$$v_i \sim U([-v_{max}, v_{max}]^3) \qquad (9b)$$

Where $r_{i+1}$ is the BP's updated position, $r_i$ is the BP's previous position, and $v_i$ is a 3D velocity vector whose components are individually sampled from a uniform distribution bounded by the maximum directional velocity. Pulmonary blood velocity varies significantly according to the diameter of the vessel, with studies finding average speeds of 18-22 cm/s [41,42]. To reduce the likelihood of BPs escaping the lung volume during random walks, $v_{max}$ was set to a conservative 20 cm/s. At each timestep, each BP's dose is determined by the partial dose at its position within the lung segmentation and added to its accumulated dose. Sixty seconds of blood circulation time was added between beams to represent gantry rotation and beam preparation time. While the length of the interval between beams was chosen arbitrarily, it has minimal impact. Shin et al. found no significant differences between blood dose histograms when varying this interval between thirty, sixty, and ninety seconds [43]. After accumulating blood dose for all beams, the resulting BP dose histogram is used to generate a bDVH.

## 2.5 Lymphocyte Survival Modeling

This study compared the radiotoxicity to CL between IMPT and pRF plans with lymphocyte radiosensitivity models. A traditional linear-quadratic (LQ) model [39] and a newer saturation model [13] that better fits recent lymphocyte survival data were used to estimate lymphocyte survival fractions (SF):

$$SF_{LQ} = \exp(-\alpha D - \beta D^2) \quad (10)$$

$$SF_S = \exp(\ln(SF_{sat})(1 - \exp(-\mu D))) \quad (11)$$

Where $D$ is dose, $\alpha$ and $\beta$ are the parameters of the LQ model, and $SF_{sat}$ and $\mu$ are the parameters of the saturation model. $SF_{sat}$ represents the SF value where SF is stable and stops decreasing with increasing dose, and $\mu$ represents the rate of increase in DNA damage with increasing dose. Saturation model parameters were derived from in vitro human lymphocyte survival data from Heylmann et al. [44]. Survival was evaluated for stimulated and unstimulated CD4 and CD8 lymphocyte subpopulations. Because the distribution of subpopulations is different for each patient, their survival was assessed individually. Radiosensitivity model parameters across the different subpopulations are shown in Table 1.

**Table 1:** Lymphocyte radiosensitivity model parameters for stimulated/unstimulated CD4 and CD8 lymphocytes. LQ parameters from Nakamura et al. only contain values for stimulated lymphocytes. Saturation parameters were calculated by Pham et al. to fit in vitro human lymphocyte survival data gathered from Heylmann et al.

| Parameter | CD4 T-Cells Stimulated/Unstimulated | CD8 T-Cells Stimulated/Unstimulated |
|---|---|---|
| $\alpha$ (Gy$^{-1}$) | 0.29 / N/A | 0.18 / N/A |
| $\beta$ (Gy$^{-2}$) | 0.14 / N/A | 0.14 / N/A |
| SF$_{sat}$ | 0.11 / 0.48 | 0.81 / 0.54 |
| $\mu$ (Gy$^{-1}$) | 0.06 / 2.67 | 1.13 / 3.79 |

bDVHs were used to estimate CL survival after a single treatment fraction with Equation 12:

$$SF = \sum_i V_i SF(d_i) \quad (12)$$

where $V_i$ is the partial blood volume receiving dose $d_i$. CL survival over the entire treatment course was computed assuming all sublethal damage was repaired between fractions:

$$SF_{tot} = \prod_j SF_j \quad (13)$$

where $SF_j$ is CL survival for fraction j. A Wilcoxon signed-rank test was performed to assess the statistical significance of survival differences.

## 2.6 Patient Study

A retrospective analysis was performed to evaluate differences in immune sparing between IMPT and pRF plans on 10 proton lung SBRT patients previously treated at our institution. Patients were prescribed 5x10 Gy (RBE) and treated with free breathing or breath-hold (BH) IMPT with 3-5 beams. Patients were robustly optimized with uniform 5 mm setup uncertainty and range uncertainties of ±3.50% for tumors near the chest wall and ±5.00% for tumors located near the center of the lung. Patient ages ranged from 55 to 89 years with a median of 75 years, and GTV volumes ranged from 0.37 to 15.93 cm$^3$ with a median of 1.65 cm$^3$

To standardize plan comparison, BH plans were used for all patients. For patients originally treated with free breathing, a BH IMPT plan was created for comparison by changing the target to the GTV and reoptimizing under plan objectives and constraints similar to the original plan. BH IMPT plans were then used to generate pRF plans for all patients. While many IMPT lung patients require free breathing plans, pRF's reduced delivery time enables more patients to qualify for BH. pRF plans were optimized to match the GTV D95% of their corresponding IMPT plan while meeting all clinical goals: lung $D_{mean}$ < 8 Gy, heart max dose < 38 Gy, heart V32 Gy < 15 cm$^3$, spinal cord max dose < 15 Gy, rib max dose < 57 Gy.

Blood dose accumulation was performed for all patients at each fraction to calculate bDVHs in three delivery scenarios: IMPT, pRF at conventional dose rates (pRF$_{CONV}$), and the same pRF plan simulated at FLASH dose rates ($pRF_{FLASH}$). A separate blood circulation simulation was performed at each fraction. bDVHs were compared between planning strategies and used to estimate CL survival for CD4 and CD8 subpopulations with Equations 12-13.

# 3. Results

Table 2 compares GTV D95%, lung-GTV $D_{mean}$, delivery time, irradiated blood volume, bDVH thresholds, and CL survival between IMPT, $pRF_{CONV}$, and $pRF_{FLASH}$. Across the cohort, pRF plans maintained the same GTV D95% as their corresponding IMPT plans. Because pRF plans had poorer energy modulation than IMPT, they had a broaderdose penumbra and increased low-dose spill, resulting in a modest increase in lungs-GTV $D_{mean}$ (mean increase 0.31 Gy, range: 0.01-0.39 Gy).

$pRF_{CONV}$/$pRF_{FLASH}$ substantially reduced delivery time (mean reductions: 85.25/99.91%, ranges: 75.44-93.56/99.86-99.9%, respectively) and the irradiated blood volume per fraction (mean reductions: 52.88/81.26%, ranges: 32.34-87.61/70.16-97.51%). After the full treatment course, $pRF_{CONV}$/$pRF_{FLASH}$ reduced irradiated blood volume (mean reductions: 7.60/40.66%, ranges: 1.55-32.45/25.94-73.52%, respectively), and IMPT irradiated nearly the entire blood volume.

**Table 2:** Plan, bDVH, and CL survival comparison across the entire patient cohort. Values are depicted as (mean, range). Statistically significant survival differences are shown in bold.

| | Single Fraction | | | Full Course | | |
|---|---|---|---|---|---|---|
| **Parameter (Mean, Range)** | **IMPT** | **$pRF_{CONV}$** | **$pRF_{FLASH}$** | **IMPT** | **$pRF_{CONV}$** | **$pRF_{FLASH}$** |
| **CTV D95% (Gy)** | | (11.38, 0.31) | | | (56.92, 1.56) | |
| **Lung-GTV $D_{mean}$ (Gy)** | (0.29, 0.21) | (0.35, 0.19) | (0.35, 0.19) | (1.44, 1.03) | (1.74, 0.95) | (1.74, 0.95) |
| **Delivery Time (s)** | (147.53, 129.21) | (19.87, 27.33) | (0.12, 0.09) | (737.65, 646.05) | (99.35, 136.65) | (0.60, 0.45) |
| **Irradiated Blood Volume (%)** | (73.90, 41.85) | (35.71, 41.55) | (14.51, 19.9) | (99.68, 1.14) | (92.16, 16.69) | (59.22, 27.49) |
| **Blood Volume at Dose (%)** | | | | | | |
| V5 cGy | (10.97, 11.83) | (3.65, 5.56) | (2.08, 3.00) | (52.65, 33.50) | (26.25, 33.56) | (13.24, 16.11) |
| V10 cGy | (8.26, 8.53) | (3.02, 4.67) | (1.72, 2.52) | (43.82, 31.12) | (21.47, 27.96) | (10.76, 13.22) |
| V20 cGy | (5.13, 3.91) | (2.44, 3.83) | (1.46, 2.18) | (32.22, 24.58) | (16.43, 21.52) | (9.18, 11.61) |
| V30 cGy | (3.28, 2.55) | (2.09, 3.34) | (1.31, 1.98) | (24.17, 21.08) | (13.39, 17.70) | (8.36, 10.73) |
| V40 cGy | (2.14, 2.01) | (1.84, 2.99) | (1.22, 1.85) | (18.34, 17.84) | (11.25, 15.15) | (7.76, 10.09) |
| V50 cGy | (1.44, 1.76) | (1.63, 2.69) | (1.13, 1.74) | (14.18, 15.36) | (9.69, 13.46) | (7.27, 9.56) |
| **Saturation CL Survival Fraction (%)** | | | | | | |
| Stimulated CD4 T-Cells | (99.61, 0.24) | (99.66, 0.58) | (99.70, 0.48) | (98.08, 1.18) | (98.35, 2.78) | (98.45, 2.52) |
| Unstimulated CD4 T-Cells | (96.87, 2.83) | (98.66, 2.07) | (98.97, 1.81) | **(85.42, 11.96)** | **(93.32, 10.24)** | **(95.03, 8.49)** |
| Stimulated CD8 T-Cells | (99.44, 0.44) | (99.69, 0.49) | (99.72, 0.48) | (97.25, 2.02) | (98.39, 2.54) | (98.66, 2.33) |
| Unstimulated CD8 T-Cells | (96.74, 3.06) | (98.69, 2.01) | (99.01, 1.73) | **(84.84, 12.87)** | **(93.46, 9.94)** | **(95.23, 8.12)** |
| **LQ CL Survival Fraction(%)** | | | | | | |
| Stimulated CD4 T-Cells | (99.02, 0.57) | (99.15, 1.40) | (99.14, 1.69) | (97.58, 1.92) | (97.91, 3.41) | (98.35, 2.78) |
| Stimulated CD8 T-Cells | (99.31, 0.43) | (99.30, 1.15) | (99.28, 1.47) | (97.16, 1.94) | (98.02, 3.18) | (98.11, 3.30) |

Figure 3 shows pRF and IMPT plans for a single patient along with their corresponding single fraction and full course bDVHs. Single fraction bDVH differences were most pronounced in the low dose region. Compared with IMPT, $pRF_{CONV}/pRF_{FLASH}$ reduced mean V5cGy to V30cGy by 7.32/8.89% to 1.19/1.97%, respectively. Smaller differences were observed at higher doses. Relative to IMPT, $pRF_{CONV}/pRF_{FLASH}$ reduced V40cGy and V50cGy by 0.30/0.92% and0.19/0.50%, respectively. Full course bDVH differences were similarly distributed. Compared to IMPT, $pRF_{CONV}/pRF_{FLASH}$ reduced meanV5cGy to V50cGy by 26.40/39.41% to 4.49/6.91%.

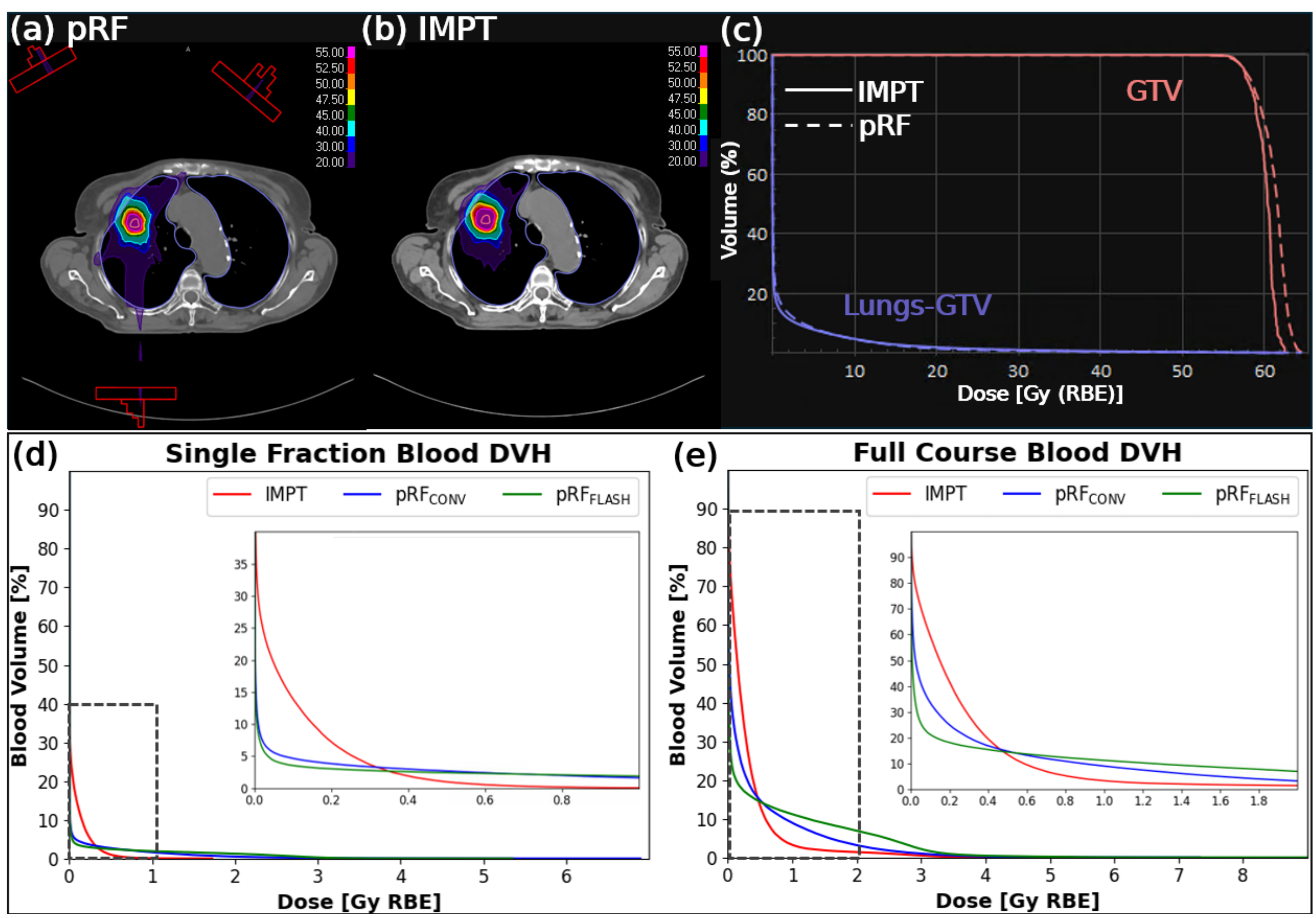

**Figure 3: (a)** 3-beam pRF plan from a patient within our cohort. pRF contours are shown in red. **(b)** Corresponding IMPT plan on the same patient. **(c)** DVH comparison between pRF and IMPT. pRF plans achieve the same GTV D95% as IMPT while slightly increasing GTV max dose and lungs-GTV dose. **(d)** Single fraction bDVH for IMPT, $pRF_{CONV}$, and $pRF_{FLASH}$. Compared to IMPT, $pRF_{CONV}/pRF_{FLASH}$ lowered irradiated blood volume at low doses, were approximately equal at approximately 35cGy, and then increased blood volume at higher doses. **(e)** Full course bDVH for IMPT, $pRF_{CONV}$, and $pRF_{FLASH}$. Relative to IMPT, $pRF_{CONV}/pRF_{FLASH}$

lowered irradiated blood volume at low doses, were approximately equal near 45cGy, and then increased blood volume at higher doses.

Figure 4 depicts survival spectra for a single patient for stimulated and unstimulated CD4/CD8 subpopulations after completing treatment. pRF plans improved CL survival at higher volume fractions while worsening CL survival at lower volume fractions. LQ survival spectra followed a sigmoid shape. Saturation survival spectra followed a sigmoid shape at higher volume fractions and converged to $SF_{sat}$ at lower volume fractions. The slope of all survival spectra increased with dose rate.

Across the patient cohort $pRF_{CONV}$ and $pRF_{FLASH}$ improved survival for all CL subpopulations and radiosensitivity models. Using the LQ model, $pRF_{CONV}$ increased mean SF by 0.33/0.86% for stimulated CD4/CD8 lymphocytes, respectively, with corresponding $pRF_{FLASH}$ increases of 0.77/0.95%. Using the saturation model, $pRF_{CONV}$ increased mean SF by 0.27/1.14% ($p = 0.32/0.20$) for stimulated CD4/CD8 lymphocytes, respectively, and by 7.90/8.62% ($p = 0.04/0.02$) in the unstimulated subpopulations, with corresponding $pRF_{FLASH}$ increases of 0.37/1.41% ($p = 0.23/0.13$) for stimulated and 9.61/10.39% ($p = 0.02/0.02$) for unstimulated subpopulations.

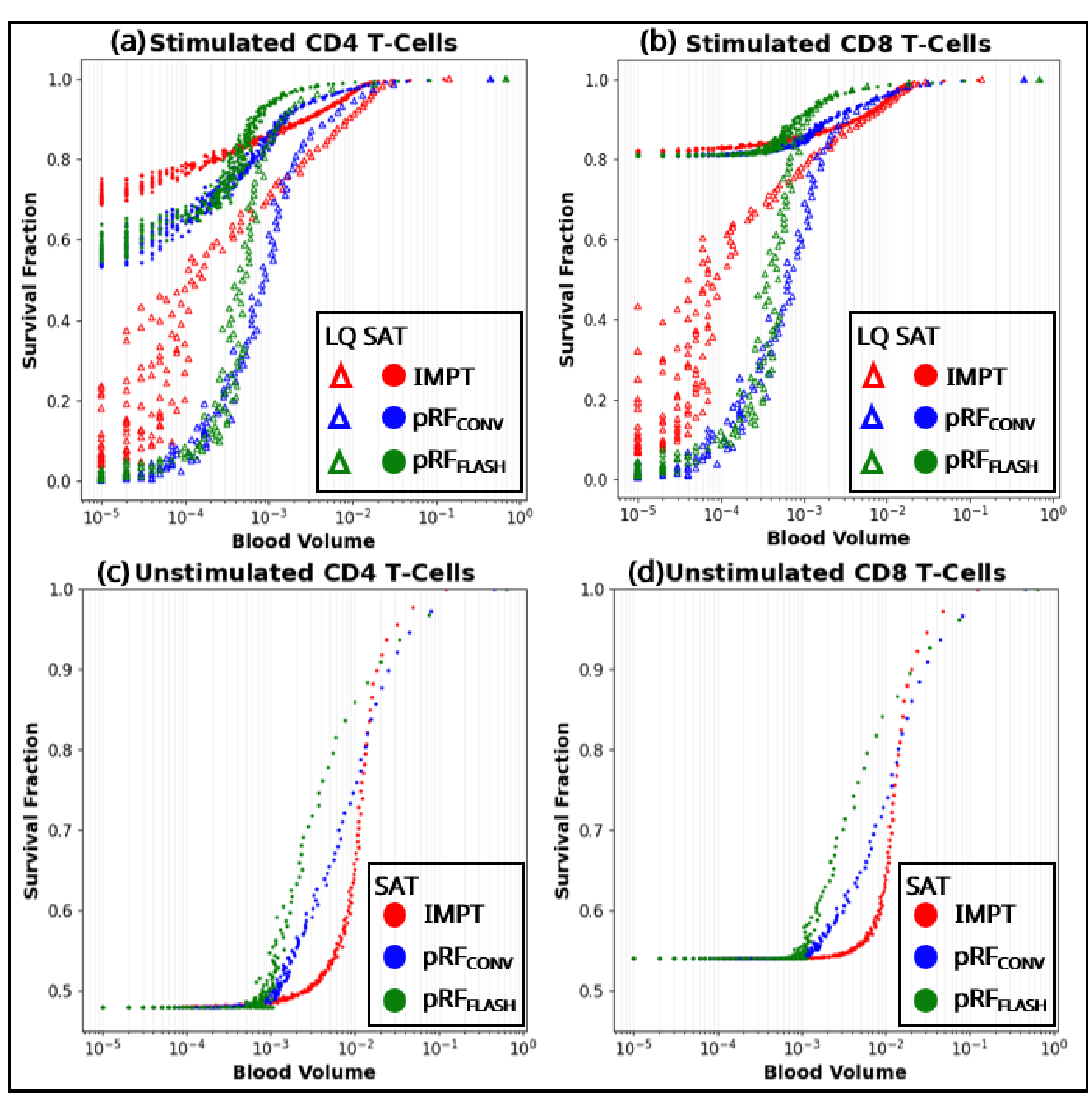

**Figure 4:** Single patient lymphocyte survival spectra after completing treatment across all planning strategies for **(a)-(b)** Stimulated CD4 and CD8 lymphocytes. Using the linear–quadratic (LQ) model, IMPT killed 3.65/2.65% of CD4/CD8 lymphocytes, respectively, $pRF_{CONV}$ killed 0.16/0.13%, and $pRF_{FLASH}$ killed 0.09/0.07%. Using the saturation model, IMPT killed 1.43/1.97%, $pRF_{CONV}$ killed 0.09/0.11%, and $pRF_{FLASH}$ killed 0.03/0.06%. **(c)-(d)** Unstimulated CD4 and CD8 lymphocytes. Using the saturation model, IMPT killed 10.44/10.86% of CD4/CD8 cells, respectively, $pRF_{CONV}$ killed 0.37/0.41%, and $pRF_{FLASH}$ killed 0.21/0.27%.

# 4. Discussion

This study evaluated patient-specific pRF delivery as a potential immune sparing strategy for proton lung SBRT. Compared with conventional IMPT, pRF plans preserved target coverage while substantially reducing delivery time and low-dose exposure of circulating blood. The modeled survival benefit was observed at both conventional and FLASH dose rates and was most pronounced for radiosensitive unstimulated CD4/CD8 lymphocytes. Importantly, this benefit was not caused by a uniform reduction in blood dose across all dose levels. Rather, pRF delivery reduced the volume of blood receiving low doses while increasing the smaller high-dose tail. This tradeoff should be considered when interpreting the biological significance of pRF delivery and when extending this approach to other treatment sites.

The clinical motivation for this work is supported by growing evidence that radiation dose to CL is associated with RIL and clinical outcomes. Early EDIC-based studies linked higher immune dose with severe lymphopenia and survival after chemoradiotherapy [34], and subsequent dynamic blood-flow models have further refined estimation of blood and lymphocyte dose during radiotherapy [35-38]. The potential advantage of pRF delivery is distinct from conventional proton dose sparing. In addition to reducing integral dose through proton beam arrangements, pRF delivery shortens the time during which circulating blood traverses the irradiated lung volume.

A key methodological contribution of this study is the integration of HEDOS-based blood circulation with a temporal simulation of PBS delivery. Prior HEDOS studies generally accumulated blood dose using organ DVHs or constant dose-rate assumptions, whereas real PBS delivery is temporally heterogeneous because spot delivery is interrupted by spot scanning and energy-layer switching. Modeling this timing structure is particularly important for comparing IMPT with pRF delivery, because the elimination of energy-layer switching is a major contributor to the reduced treatment time of pRF plans. Continuous scanning between closely spaced spots was not explicitly modeled. Because the impact of this simplification depends on spot spacing, spot ordering, and machine-specific beam-on behavior during scanning, future work should validate the temporal model directly against delivery log files and, when possible, include continuous scanning in blood dose accumulation.

$pRF_{FLASH}$ provided only a modest additional improvement relative to $pRF_{CONV}$ despite a much higher assumed dose rate. Once delivery becomes sufficiently short, only a limited additional volume of blood can enter the irradiated lung during treatment. Therefore, further increases in dose rate produce diminishing reductions in irradiated blood volume. This apparent saturation suggests that a substantial component of the pRF immune-sparing effect may be achieved through ultrafast single-energy delivery even at conventional dose rates, rather than depending entirely on FLASH. The FLASH simulations assumed a constant 500 nA nozzle current across all energies. This should be interpreted as a hypothetical FLASH-delivery scenario rather than a direct representation of clinical delivery at all energies.

The selection of lymphocyte survival model is important for interpreting these results. We included the LQ model because the Nakamura et al. parameters provide a conventional benchmark for stimulated CD4/CD8 lymphocyte radiosensitivity [39]. However, the conventional LQ model may not fully capture the nonlinear dose response reported in lymphocyte apoptosis and survival assays. For this reason, the saturation model from Pham et al. was used as the primary model [13]. Pham et al. systematically compared several dose-response models for lymphocyte data and reported that the saturation model provided a better fit than the LQ model, including lower residual error and improved Akaike information criterion (AIC) performance, particularly for unstimulated lymphocyte datasets [13]. This is relevant to the present study because the largest pRF-associated differences occurred in the low-dose region of the bDVH and in unstimulated lymphocyte subpopulations, where model selection can meaningfully affect the estimated survival benefit.

This study used a saturation model from Pham et al.  that was fit to in vitro human lymphocyte data from Heylmann et al. [44]. This model was chosen because it allowed the evaluation of survival in stimulated and unstimulated CD4/CD8 lymphocyte subpopulations. Stimulated lymphocytes are more radioresistant, whereas resting or unstimulated lymphocytes are more radiosensitive. Accordingly, modeled survival gains were not significantly different for stimulated lymphocytes but were significantly larger for unstimulated lymphocytes. These results suggest that pRF delivery preferentially spares the most radiosensitive CL, which make up >95% of the CL population [46,47].

The current survival estimates should be interpreted as relative biological comparisons rather than direct predictions of clinical lymphocyte counts. Both the LQ and saturation models used in this study were derived from in vitro lymphocyte irradiation data. Such models do not account for in vivo microenvironmental factors, lymphocyte trafficking between blood and lymphoid organs, immune-cell redistribution, bone marrow and lymphoid-organ dose, or patient-specific immune recovery. However, these limitations do not necessarily imply that the modeled pRF benefit is overestimated. In fact, several lines of evidence suggest that the clinical effect of reducing low-dose CL exposure could be larger in patients than predicted by in vitro survival models. Lymphocytes have been reported to be more radiosensitive in vivo than in vitro, potentially due to plasma-mediated nitric oxide signaling, impaired DNA repair, and enhanced apoptosis after whole-body irradiation [47]. In addition, patients receiving concurrent chemotherapy or immunotherapy may have reduced lymphocyte reserve and delayed immune recovery [48,49], making reductions in circulating blood dose more consequential. Therefore, the absolute SF values reported here should be viewed cautiously, but the relative improvement with pRF delivery may be clinically meaningful and warrants prospective validation.

Several additional limitations should be acknowledged. This was a retrospective planning and modeling study of 10 lung SBRT patients, and the bDVH and CL survival endpoints were not correlated with measured absolute lymphocyte counts or lymphocyte subsets. Patient-specific blood volume, cardiac output, and intrapulmonary blood-flow patterns were not modeled; instead, reference values and a simplified random-walk model were used. The study also standardized patients to breath-hold plans for comparison, which does not represent all clinical lung SBRT workflows. Future studies should incorporate patient-specific hemodynamics when available, delivery-log-based timing, measured absolute lymphocyte counts, and clinical endpoints such as lymphocyte nadir, recovery kinetics, infection risk, immunotherapy response, and survival.

Taken together, these findings support pRF delivery as a promising strategy for immune sparing proton lung SBRT. The principal advantage appears to arise from reducing the duration of blood irradiation and lowering low dose bDVH exposure, with additional but saturating gains at theoretical FLASH dose rates. Because the survival models are based on in vitro data and because clinical RIL is influenced by multiple patient and treatment-related factors, the present results should be considered hypothesis-generating. Prospective clinical studies are needed to determine whether the modeled improvement in CL survival translates into measurable reductions in RIL and improved immune-related outcomes.

# 5. Conclusions

This study evaluated the potential for pRF delivery to enhance immune preservation in proton lung SBRT by comparing clinically deliverable pRF plans with conventional IMPT. By substantially reducing treatment time through the elimination of ELS, pRF plans decreased low dose blood exposure while slightly increasing the fraction receiving high dose, improving CL survival at both conventional and FLASH dose rates. These effects were most pronounced for radiosensitive, unstimulated lymphocyte populations, suggesting that even modest reductions in CL dose may have a meaningful biological impact. Our findings suggest that pRFs represent a promising strategy to mitigate RIL and improve immune outcomes in proton lung SBRT.